\documentclass{aa}
\usepackage{graphicx}

\begin{document}

   \title{Star clusters in the Carina complex:\\ $UBVRI$
 photometry of NGC~3114, Collinder~228 and vdB-Hagen~99\thanks{Based 
on observations taken at ESO La Silla. Data are available in electronic 
form at the CDS via anonymous ftp to cdsarc.u-strasbg.fr (130.79.128.5)}}

   \author{G. Carraro\inst{1,2} and F. Patat\inst{2}    
		  }

   \offprints{G. Carraro ({\tt carraro@pd.astro.it})}

   \institute{Dipartimento di Astronomia, Universit\'a di Padova,
	vicolo dell'Osservatorio 5, I-35122, Padova, Italy
        \and
	European Southern Observatory, 
 Karl-Schwartzschild-Str 2, D-85748 Garching b. M\"unchen, Germany\\
 e-mail: {\tt 
carraro@pd.astro.it,fpatat@eso.org}
             }

   \date{Received ; accepted}

\abstract{
In this paper we  present and analyze CCD  $UBVRI$ photometry 
in the region of the three
young open clusters NGC~3114, Collinder~228, and vdB-Hagen~99,
located in the Carina spiral feature.\\
NGC~3114 lies
in the outskirts of the Carina nebula. We found 7 star members
in a severely contaminated field, and obtain a distance of 950 pc
and an age less than  $3 \times 10^{8}$~yrs.\\
Collinder~228 is a younger cluster  (8$\times 10^{6}$~yrs), located
in front of the Carina nebula complex, for which we  identify 11 new
members and suggest that 30\% of the stars are probably binaries.\\
As for vdB-Hagen~99, we add 4 new members, confirming  that it 
is a nearby cluster located 
at 500 pc from the Sun and  projected toward the direction of the Carina 
spiral arm.
}
\authorrunning{Carraro \& Patat}

\maketitle

%

\section{Introduction}
This paper continues a series dedicated at presenting
CCD $UBVRI$ photometry for all the known and/or suspected
open clusters in the Carina complex (Janes et al 1988).
According to Feinstein (1995) the region around $\eta$~Carinae
contains 14 open clusters, some of which
still remain very poorly studied. For most of them
no CCD photometry was available when we started our survey.\\
In Carraro et al (2001) we discussed NGC~3324 and Loden~165,
concluding that the latter has probably no relation with the 
Carina Complex, being much older and closer to the Sun than the bulk
of the other clusters.\\
In Patat \& Carraro (2001) we studied Bochum~9, 10 and 11,
suggesting that Bochum~9 is a doubtful object, while
Bochum~10 and 11 are two young and poorly populated
open clusters.\\
In this work we present results for NGC~3114, Collinder~228 and 
vdB-Hagen~99, for which to our knowledge no multicolor
CCD photometry is available.\\
The aim is to provide accurate photometry for all
these clusters, in order to derive
precise age estimates. This is important to infer
global properties and
to study the clusters formation history
in the very interesting Carina region.\\
The layout of the paper is as follows: Sect.~2 presents very briefly
the data acquisition and reduction. In Sect.~3 we discuss the open
cluster NGC~3114.
Sect.~4 is dedicated to Collinder~228,
whereas Sect.~5 deals with vdB-Hagen~99.
Our conclusions
are summarized in Sect.~6.

\begin{table}
\caption{Basic parameters of the observed objects.
Coordinates are for J2000.0 equinox.}
\begin{tabular}{lcccc}
\hline
\hline
\multicolumn{1}{l}{Name} &
\multicolumn{1}{c}{$\alpha$}  &
\multicolumn{1}{c}{$\delta$}  &
\multicolumn{1}{c}{$l$} &
\multicolumn{1}{c}{$b$} \\
\hline
& $hh:mm:ss$ & $^{o}$~:~$^{\prime}$~:~$^{\prime\prime}$ & $^{o}$ & $^{o}$ \\
\hline
NGC~3114          & 10:02:42.7 & -60:06:32.1 & 283.34  & -3.83\\
Collinder~228     & 10:43:01.3 & -60:00:44.8 & 287.52  & -1.04\\
vdB-Hagen~99      & 10:37:54.2 & -59:11:37.1 & 286.56  & -0.63\\
\hline\hline
\end{tabular}
\end{table}

\begin{table*}
\tabcolsep 0.10truecm
\caption{Journal of observations of NGC~3114 (April 13, 1996),
Collinder~228 (April 14 , 1996), and vdB-Hagen~99 (April 15, 1996).}
\begin{tabular}{cccccccccccccccc} \hline
      & \multicolumn{3}{c}{NGC~3114}         & \multicolumn{3}{c}{Collinder~228}          & \multicolumn{9}{c}{vdB-Hagen~99} \\
Field & Filter & Exp. Time & Seeing   & Field       & Filter & Exp. Time & seeing & Field & Filter & Exp. Time & seeing & Field &Filter & Exp. Time & seeing  \\
      &        & (sec)     & ($\prime\prime$)&  &      &  (sec)    & ($\prime\prime$) & &       &  (sec)    & ($\prime\prime$) & & & (sec)    & ($\prime\prime$)\\
      &        &           &                 &        &           &  & & &                 &        &           & \\  
 $\#1$   & U & 1200 &  2.0 & $\#1$  &  &      &      &$\#1$ &  &     &    & $\#3$  &   &     &    \\
	 & U &  120 &  1.8 &        &U &   60 &  2.0 & &U & 180 & 2.3&        & U & 300 & 1.7\\
	 & B &   10 &  1.7 &        &B &   10 &  2.0 & &  &     &    &        & B &  30 & 1.6\\
	 & B &  900 &  1.7 &        &B &  300 &  1.6 & &B &  30 & 2.1&        & B & 300 & 1.5\\
	 & V &   20 &  1.5 &        &V &    3 &  1.7 & &  &     &    &        & V &  20 & 1.6\\
	 & V &  300 &  1.6 &        &V &  120 &  1.8 & &V &  15 & 2.0&        & V & 120 & 1.6\\
	 & R &   10 &  1.5 &        &R &    3 &  1.7 & &  &     &    &        & R &   5 & 1.6\\
	 & R &  180 &  1.4 &        &R &   60 &  1.8 & &R &  10 & 1.9&        & R & 120 & 1.6\\
	 & I &    5 &  1.3 &        &I &    5 &  1.8 & &  &     &    &        & I &   5 & 1.5\\
	 & I &  300 &  1.3 &        &I &  120 &  1.6 & &I &  20 & 1.7&        & I & 120 & 1.6\\
 $\#2$   & U &   60 &  1.5 & $\#2$  &  &      &      &$\#2$ &U & 300 & 2.4& &&&\\
	 & U & 1200 &  1.6 &        &U &   60 &  1.6 & &U &  60 & 2.0& &&&\\
	 & B &   10 &  1.4 &        &B &   10 &  1.6 & &B & 300 & 2.3& &&&\\
	 & B &  900 &  1.4 &        &B &  300 &  1.8 & &B &  30 & 2.3& &&&\\
	 & V &   10 &  1.4 &        &V &    3 &  2.0 & &V & 120 & 2.1& &&&\\
	 & V &  300 &  1.5 &        &V &  120 &  2.0 & &V &  10 & 2.0& &&&\\
	 & R &    5 &  1.4 &        &R &    3 &  1.8 & &R &   5 & 1.8& &&&\\
	 & R &  180 &  1.2 &        &R &   60 &  1.9 & &R & 120 & 1.8& &&&\\
	 & I &   10 &  1.2 &        &I &    5 &  1.9 & &I & 120 & 1.8& &&&\\
	 & I &  300 &  1.3 &        &I &  120 &  2.0 & &I &   5 & 1.8& &&&\\ 
\hline
\end{tabular}
\end{table*}

\section{Observations and Data Reduction}
Observations were conducted at La Silla on April 13-16, 1996,
with the 0.92m ESO--Dutch telescope. 
The observations strategy, the data
reduction, the error analysis and a comparison between CCD and
photoelectric photometry   
have been presented in Patat \& Carraro (2001), which the reader is
referred to for any detail. Finally, all the data are available
upon request to the authors.

\begin{figure}
\centering
\includegraphics[width=9cm,height=9cm]{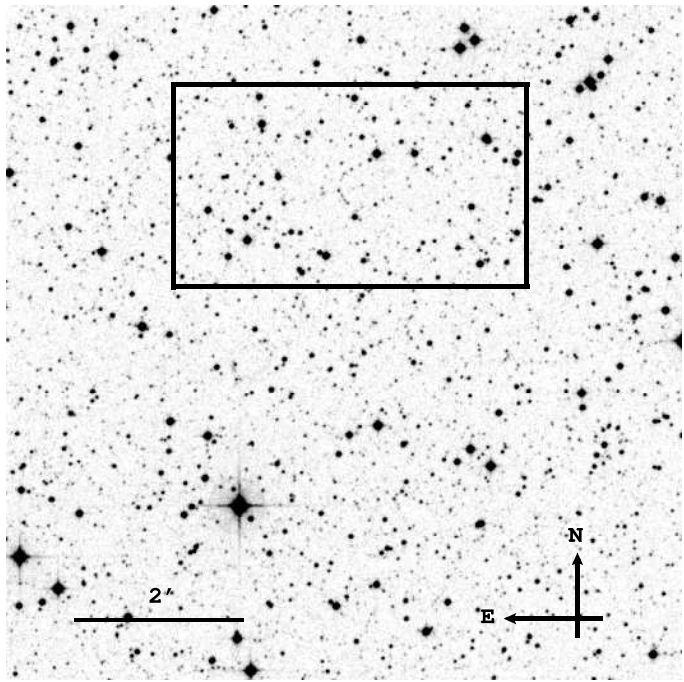}
\caption{DSS map of a region around NGC~3114. The box confines the field
covered by our photometry.}
\end{figure}

\begin{figure}
\centering
\includegraphics[width=9cm,height=9cm]{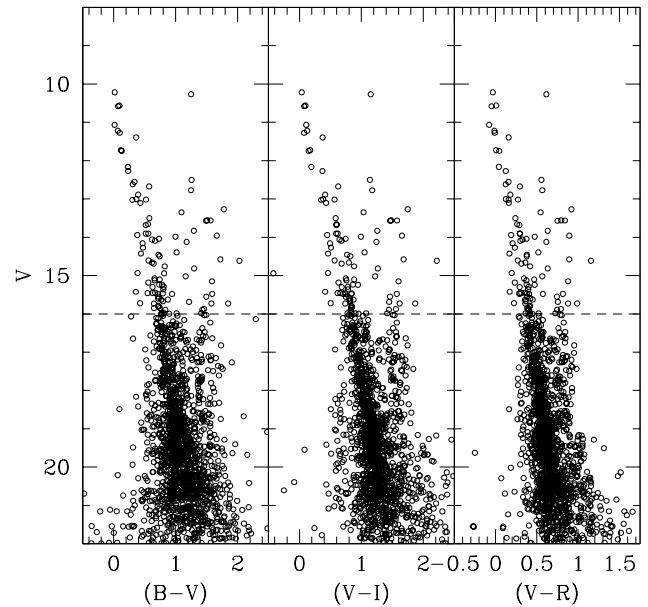}
\caption{CMDs for all stars in the region of NGC~3114.
The dashed line indicates the limiting magnitude reached
by Sagar \& Sharpless (1991).}
\end{figure}

\section{The open cluster NGC~3114}
NGC~3114 is a sparse open cluster projected onto the outskirts
of the Carina complex,
in a fairly rich Milky Way field. Its membership to
the Carina complex is actually not clear.
It is a difficult object to study,
due to the heavy contamination of Galactic disk field stars
which does not allow one to unambiguously separate possible members
and define the cluster size.

\subsection{Previous results}
NGC~3114 was studied several times in the past.
The first investigation was performed by Jankowitz
and McCosh (1963), who  obtained photographic $UBV$ photometry for 171 stars
and photoelectric $UBV$ photometry of 52 stars down to $V$~=~13 mag.
They estimated that the cluster is 910 pc distant from the Sun,
has a mean visual extinction E$(B-V)$~=~0.27, and an age between
$6 \times 10^{7}$ and $2 \times 10^{8}$~yrs.\\
Afterward Schneider \& Weiss (1988) got Str\"omgrem photometry of 122
stars down to $V$~=~12 mag. This study  strongly revises
the cluster reddening, which the authors suggested to be
E$(B-V)$~=~0.03.\\
More recently, Sagar \& Sharpless (1991) enlarged the sample
of the measured stars, obtaining $BV$ CCD photometry of about 350 stars
up to $V$~=~16 in seven $3^{\prime}.6 \times 5^{\prime}.4$ regions
located quite far from the cluster center, 
where the contamination is 
expected to be more important.
By assuming the reddening estimate suggested by
Schneider \& Weiss (1988), they found a cluster distance of 940$\pm$60 pc,
in agreement with Jankowitz
and McCosh (1963), and an age of $1-2 \times 10^{8}$~yrs.\\
Finally, Clari\`a et al (1989) estimated the cluster mean chemical
abundance from UBV, DDO and Washington photometry of an handful of giant stars, 
finding that NGC~3114 is basically as metal rich as the Sun 
($[Fe/H]~=~-0.04\pm0.04$).

\begin{table}
\tabcolsep 0.20cm
\caption{Photometry of the stars in the field
of  the open cluster NGC~3114 in common with Jankowitz \& McCosh (1963).
The suffix $CP$ refers to the present study, whereas $JM$ indicates
Jankowitz \& McCosh (1963) photographic photometry.}
\begin{tabular}{cccccc}
\hline
\hline
\multicolumn{1}{c}{ID} &
\multicolumn{1}{c}{$JM$}  &
\multicolumn{1}{c}{$V_{CP}$}  &
\multicolumn{1}{c}{$(B-V)_{CP}$} &
\multicolumn{1}{c}{$V_{JM}$}  &
\multicolumn{1}{c}{$(B-V)_{JM}$} \\
\hline
 3  & 105 & 10.580  & 0.069  & 10.59 & -0.08   \\
 4  & 122 & 10.562  & 0.092  & 10.57 & -0.03   \\
 5  & 137 & 11.274  & 0.097  & 11.27 &  0.10   \\
 6  &  73 & 11.065  & 0.014  & 11.12 &  0.00   \\
 7  & 130 & 11.393  & 0.364  & 11.38 &  0.22   \\
 9  &  83 & 11.222  & 0.068  & 11.24 &  0.11   \\
10  &  76 & 11.726  & 0.118  & 11.78 &  0.02   \\
11  &  96 & 10.268  & 1.250  & 10.22 &  1.12   \\
13  & 109 & 12.272  & 0.235  & 12.12 &  0.36   \\
18  & 128 & 13.004  & 0.302  & 12.90 &  0.36   \\
\hline
\end{tabular}
\end{table}

\subsection{The present study}
We provide $UBVRI$ photometry for 2060 stars in a $3^{\prime}.3 \times
6^{\prime}.5$ region centered in NGC~3114, up  to about V~=~22. 
The region we sampled is shown in Fig.~1, where a DSS\footnote{(Digital Sky
Survey {\tt http://archive.eso.org/dss/dss})} map is presented.\\
According to Jankowitz
and McCosh (1963) the cluster should have a diameter of $32^{\prime}$,
although this estimate is rather uncertain, due to the difficulty
to isolate the cluster from the field. Anyhow, the cluster seems
to be rather extended, and our photometry covers only the central region, 
with no overlap with Sagar \& Sharpless (1991) photometry.\\
The CMDs for all the measured stars in the planes 
$V-(B-V)$, $V-(V-I)$ and $V-(V-R)$ are shown in Fig.~2. 
In the cluster center there are no stars brighter than $V$~=~10.0.
Therefore, with respect to
Sagar \& Sharpless (1991), who provided the deepest photometry before
our study,  we do not find any indication of a Red Giant (RG) 
clump (see Fig.~9a in Sagar \& Sharpless 1991).

The Main Sequence (MS) extends from $V$~=~10 up to $V$~=~22, and gets
wider at increasing magnitudes. Several causes concur to broaden
the MS: the presence of unresolved binary stars, the photometric errors
and the contamination of fore-ground and back-ground stars.\\
A probe of the  heavy contamination is the Galactic disk RG 
branch population, readily recognizable
in the almost parallel sequence 
which departs from the MS at $V$ $\approx$ 20. 
This is a common feature in the CMDs of stellar fields
in the direction of the Carina spiral arm (see Vallenari et al 2000).

We have 10 stars in common with Jankowitz \& McCosh (1963),
which are listed in Table~3. The mean differences turn out to be:

\[
V_{CP} - V_{JM} = 0.020\pm0.062 
\]

\[
(B-V)_{CP} - (B-V)_{JM} = 0.068\pm0.077 
\]

\noindent
where $CP$ indicates our photometry, and $JM$ stands for
Jankowitz \& McCosh (1963).
Taking into account the different techniques used in extracting
the photometry, the agreement is good for both magnitude and colors.\\
We do not report the difference between the $(U-B)$ colors,
since Jankowitz \& McCosh (1963) measured the color $(U_{c}-B)$,
with the filter $U$ defined in the Cousin system.
This is quite different from the standard Johnson $(U-B)$, and
the authors do not provide the Johnson color for the
stars listed in Table~3.\\

\begin{figure}
\centering
\includegraphics[width=9cm,height=9cm]{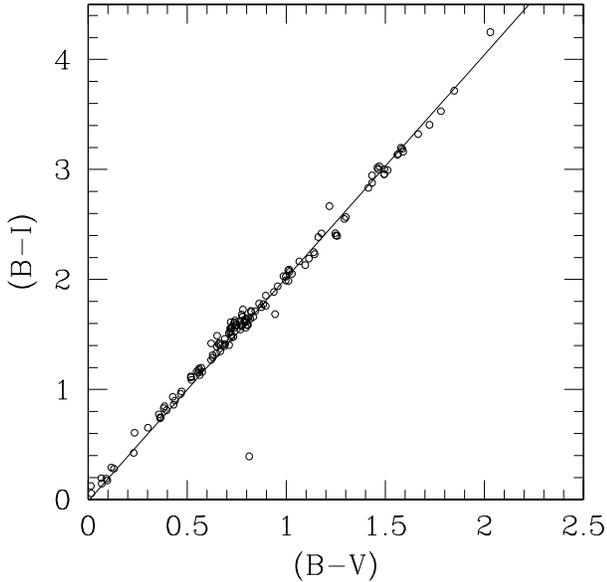}                               \caption{NGC 3114 stars brighter than $V$~=17.0 in the
$(B-I)-(B-V)$ plane.}
\label{bi3114}
\end{figure}

\begin{figure}
\centering
\includegraphics[width=9cm,height=9cm]{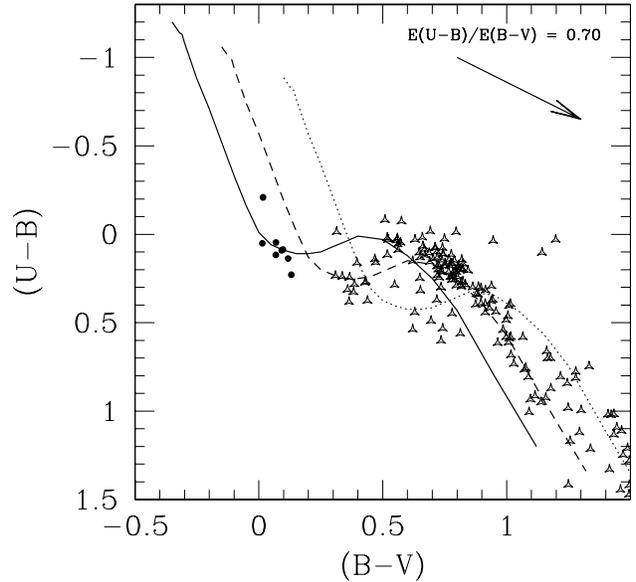}       
\caption{Two color diagram for the stars in the field of NGC~3114
with $V$ $\leq$~17.0.The 
arrow indicates the reddening vector. The solid line is the empirical
un-reddened ZAMS from Schmidt-Kaler (1982), whereas the dashed and
dotted lines are the same ZAMS, but shifted by E$(B-V)$~=~0.20
and E$(B-V)$~=~0.60, respectively.}
\end{figure}

\begin{figure}
\centering
\includegraphics[width=9cm,height=9cm]{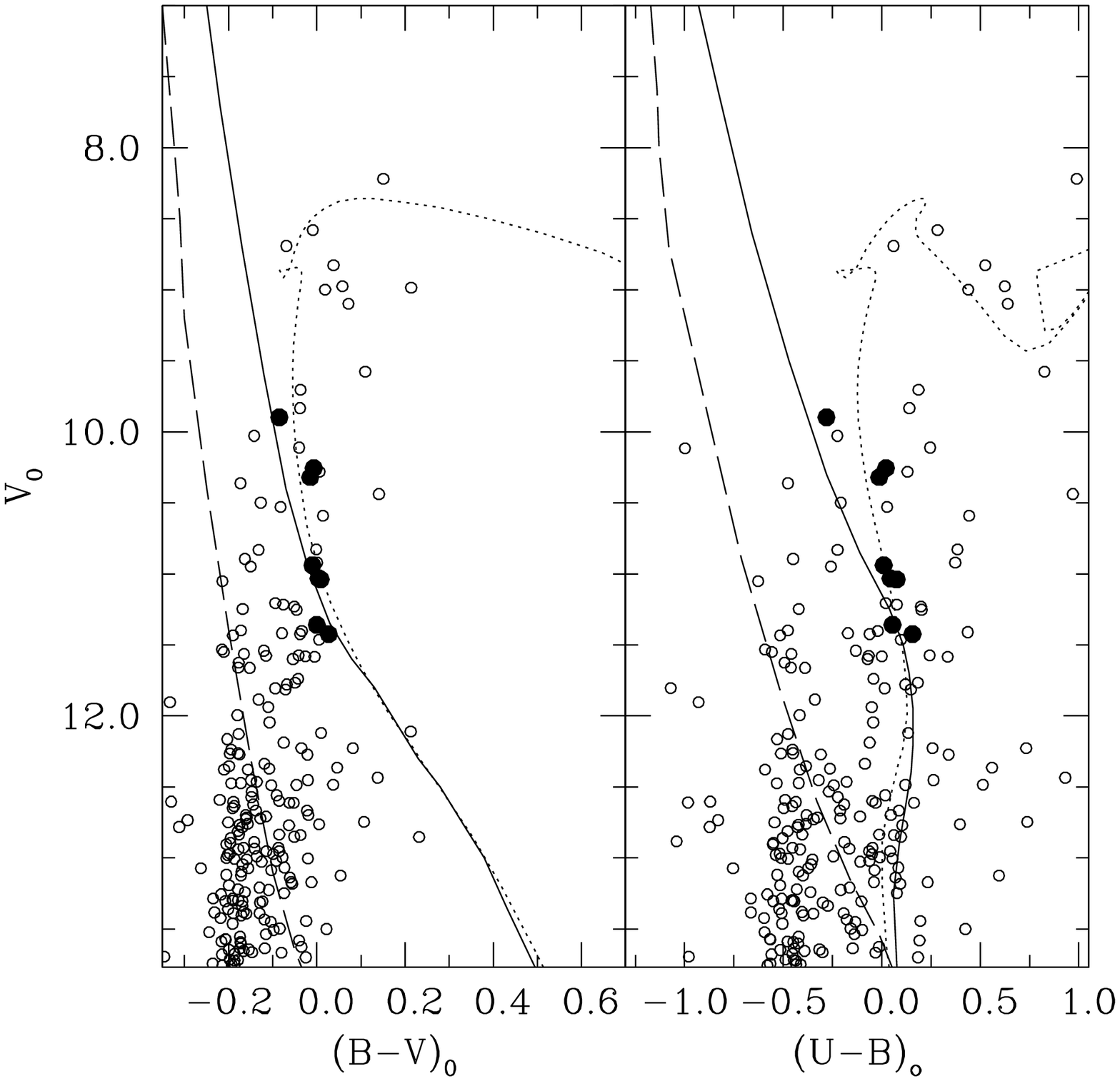}                               
\caption{Reddening corrected CMDs for the stars in the field of NGC~3114
with $V$ $\leq$ 17.0 The dashed line is a ZAMS shifted by (m$-$M)=12.50,
whereas the solid one is the same ZAMS, but shifted by (m$-$M) = 9.80.
Filled circles indicate cluster candidate members, while empty circles 
indicated field stars. Finally, the dotted line is an isochrone
for the age of 300 Myr.}
\end{figure}

\subsection{Reddening}
In order to obtain an estimate of the cluster mean reddening, we 
analyse the distribution of the stars in the $(B-I)-(B-V)$ plane,
which is shown in Fig.~\ref{bi3114}.\\
The linear fit to the main sequence in the $(B-I)-(B-V)$ plane,

\begin{equation}
(B-I) = Q + 2.25 \times (B-V)
\end{equation}

\noindent
can be expressed in terms of E$(B-V)$ for the $R_V$~=~3.1
extinction law as

\begin{equation}
E(B-V) = \frac{Q-0.014}{0.159}   ,
\label{mu2}
\end{equation}

\noindent
following the method proposed by  Munari \& Carraro (1996a,b).\\
This method provides a rough estimate of the mean reddening and,
as amply discussed in Munari \& Carraro (1996a), can be used
only for certain color ranges. In particular Eq.~\ref{mu2}
holds over the range $-0.23 \leq (B-V)_o \leq +1.30$.
A least squares  fit through the stars brighter than $V$~=17 
gives Q~=~0.041,
which, inserted in Eq.~\ref{mu2}, provides  E$(B-V)$=0.17$\pm$0.12.\\
The uncertainty is rather large, and is due to the scatter 
of the stars in this plane, which indicates the presence
of stars with different reddening, presumably a mixture
of stars belonging to the cluster and to the field.\\

\begin{table*}
\tabcolsep 0.50cm
\caption{List of the candidate members of NGC~3114 obtained in
the present study. Two of them ($\#3$ and $\#8$) are newly discovered
members. }
\begin{tabular}{cccccccc}
\hline
\hline
\multicolumn{1}{c}{ID} &
\multicolumn{1}{c}{$X$}  &
\multicolumn{1}{c}{$Y$}  &
\multicolumn{1}{c}{$V$} &
\multicolumn{1}{c}{$(B-V)$}  &
\multicolumn{1}{c}{$(U-B)$} &
\multicolumn{1}{c}{$(V-R)$} &
\multicolumn{1}{c}{$(R-I)$} \\
\hline
   2 &  289.93 &  159.96  &   10.216  &   0.017 &   -0.209 &   -0.142 &  0.070 \\
   3 &   60.20 &  129.03  &   10.580  &   0.069 &    0.046 &   -0.047 &  0.124 \\
   4 &  271.54 &  495.20  &   10.562  &   0.092 &    0.091 &    0.012 &  0.090 \\
   5 &  408.86 &  238.26  &   11.274  &   0.097 &    0.086 &   -0.011 &  0.087 \\
   6 & -384.24 &  137.45  &   11.065  &   0.014 &    0.051 &   -0.077 &  0.196 \\
   8 &  213.23 &  346.53  &   11.746  &   0.131 &    0.229 &    0.045 &  0.013 \\
   9 & -173.94 &  442.02  &   11.222  &   0.068 &    0.117 &   -0.014 &  0.118 \\
\hline
\end{tabular}
\end{table*}

\noindent
To better derive the reddening distribution and identify cluster members,
we plotted all the stars brighter than 
$V$~=~17 in the two color diagram of Fig.~4.
With filled circles we indicated stars having a common low reddening
E$(B-V)$~=~0.07$\pm$0.03. They lie very close to the
unreddened  empirical ZAMS (solid line) taken from Schmidt-Kaler (1982) .\\
Open triangles represent  all the other stars, which exhibit a much 
larger reddening.  These stars do not suffer from the same amount of 
reddening. To have an idea of the reddening of the field stars
we have overimposed the same ZAMS, but shifted by E$(B-V)$~=~0.20
(dashed line), and by E$(B-V)$~=~0.60 (dotted line), respectively.\\
In conclusion, two populations seem to exist: seven stars
sharing a common low reddening, which are presumably cluster
members, and all the other stars having larger
reddening,  which are field stars.

\subsection{Age and distance}
We estimate the age of NGC~3114 by studying the reddening
corrected CMDs (see Fig.5). In this plot filled circles
are our candidate members, whereas empty circle are field
stars. The bulk of this latter is fitted by a ZAMS
shifted by (m$-$M) = 12.50 (dashed line), basically at the distance
of the Carina spiral arm. Nevertheless, there seem to be
stars located basically at any distance between us and the Carina
spiral arm, confirming previous indications that the cluster
is heavily contaminated by stars from the Galactic disk field.\\
\noindent
The candidate members form a tight sequence,
close to a ZAMS shifted by (m-M)=9.80 (solid line).\\
In order to estimate the cluster age, we over-imposed
a solar metallicity isochrone (dotted line) from Girardi et al (2000),
for the age of $3 \times 10^{8}$ yrs. In fact, the brightest
members lie off the ZAMS, and are clearly leaving the MS.\\

\noindent
In conclusion, NGC~3114 is a rather poorly populated star cluster, heavily
contaminated by field stars. From the study of the cluster
candidate members (see Table~4), in our field we derive a reddening E$(B-V)$~=~0.07$\pm$0.03
and a distance of 920$\pm$50 pc from the Sun, in fair agreement
with previous investigations.

\begin{figure}
\centering
\includegraphics[width=9cm,height=9cm]{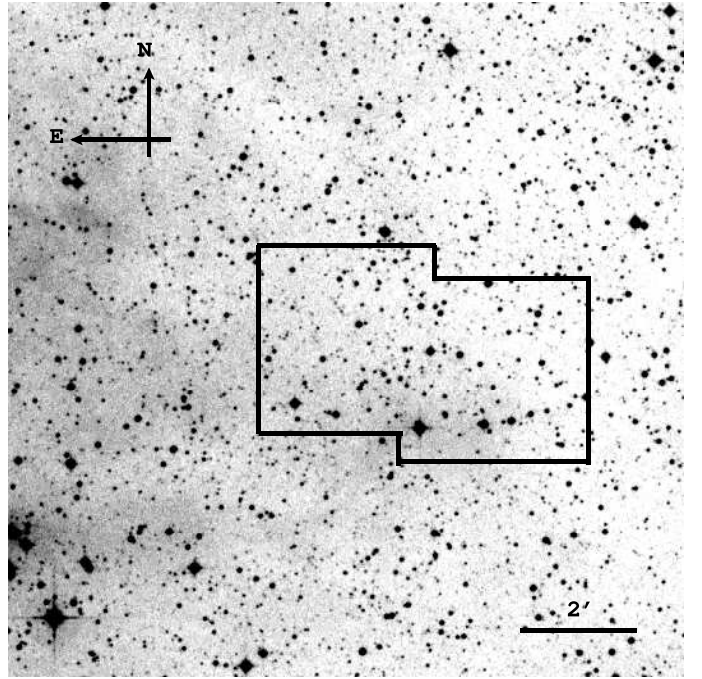}   
\caption{DSS map of a region around Collinder~228. The box confines the field covered by our photometry.}
\end{figure}

\section{The open cluster Collinder~228}
Collinder~228 was discovered by Collinder (1931) during a systematic search
of open clusters in the Milky Way, and lies between 
us and the group formed by Trumpler~14, Trumpler~16 and $\eta$~Carinae
(Smith et al 2000). Therefore we expect the cluster to be dominated
by back and foreground stars contamination.
Differential reddening is also expected, since the cluster
is surrounded by a large nebula (see Fig.~6).

\begin{figure}
\centering
\includegraphics[width=9cm,height=9cm]{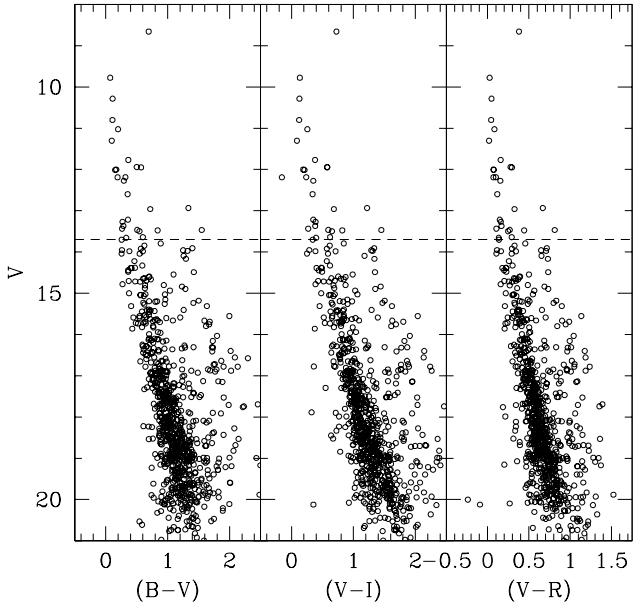}     
\caption{CMDs for all stars in the region of Collinder~228.
The dashed line indicates the limiting magnitude reached by
Feinstein et al (1976).}
\end{figure}

\subsection{Previous results}
Feinstein et al. (1976) reported on $UBV$ photoelectric photometry
of 99 stars in the region of Collinder~228. They found
that the bulk of the cluster is located in front of the complex
of Trumpler~14 and 16, at about 2.5 kpc from the Sun. 
They also pointed out that some stars in the field
of Collinder~228 might be members of that complex and hence
more distant than the cluster. They assign to Collinder~228 an
age of $5 \times 10^{6}$~yrs. While the bulk of the stars
is closer and has a mean reddening E$(B-V)$~=~0.30, 
the stars lying  beyond the cluster have a larger reddening
E$(B-V)$~$\approx$~0.50.\\
Tapia et al (1988) obtained $JHKL$ near-infrared photometry
of 200 stars in the $\eta$~Carinae region, which comprises
Trumpler~14, 15, 16 and Collinder~228 and 232.
Out of these, 45 are in the field of Collinder~228. By analyzing
the two color diagrams, the authors concluded that this cluster is 
$2.09\pm0.38$ kpc far from the Sun, and  hence closer to us than 
the bulk of $\eta$~Carinae region population.
Moreover they found that the mean reddening
is E$(B-V)$~=~0.64$\pm$0.26, much larger than the value previously reported
by Feinstein et al. (1976).\\

Finally, a radial velocity survey has been conducted by 
Levato et al (1990), who suggested that 30\% of the cluster
stars are binaries.

\begin{figure}
\centering
\includegraphics[width=9cm,height=9cm]{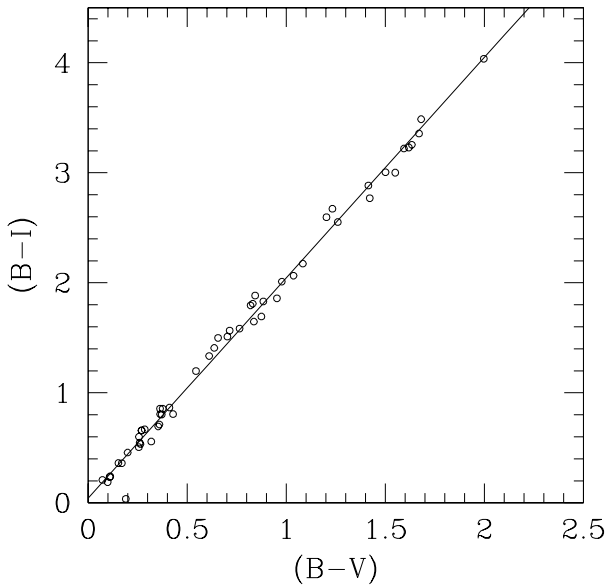} 
\caption{Collinder 228 stars brighter than $V$~=17 in the
$(B-I)-(B-V)$ plane.}
\label{bi228}
\end{figure}

\subsection{The present study}
We provide $UBVRI$ photometry for about 1100 stars in a
$3^{\prime}.3 \times 6^{\prime}.5$ region 
centered in Collinder~228, up to about $V$~=~21. The covered
region is shown in Fig.~6.

The CMDs for all the measured stars are shown in Fig.~7 in the planes
$V-(B-V)$, $V-(V-I)$ and $V-(V-R)$.
The MS extends from $V$~=~10 down to $V$~=21.
As for NGC~3114, the MS gets wider at increasing magnitude, and there is
evidence of a secondary sequence on the red side of the MS,
generated by the RG stars in the field.

\begin{table*}
\tabcolsep 0.20cm
\caption{Photometry of the stars in the field
of  the open cluster Collinder~228 in common with Feinstein et al (1976).
The suffix $CP$ refers to the present study, whereas $FMF$ indicates
Feinstein et al (1976) photometry.}
\begin{tabular}{ccccccccc}
\hline
\hline
\multicolumn{1}{c}{ID} &
\multicolumn{1}{c}{$FMF$}  &
\multicolumn{1}{c}{$Name$}  &
\multicolumn{1}{c}{$V_{CP}$}  &
\multicolumn{1}{c}{$(B-V)_{CP}$} &
\multicolumn{1}{c}{$(U-B)_{CP}$} &
\multicolumn{1}{c}{$V_{FMF}$}  &
\multicolumn{1}{c}{$(B-V)_{FMF}$} &
\multicolumn{1}{c}{$(U-B)_{FMF}$} \\
\hline
 1  & 15  & HD~305544 &  8.656   & 0.694   &  0.192  &  8.59  & 0.66  &  0.08   \\
 2  & 28  & HD~305543 &  9.778   & 0.073   & -0.768  &  9.74  & 0.05  & -0.77   \\
 3  & 29  & HD~305451 & 10.284   & 0.112   & -0.265  & 10.21  & 0.07  & -0.36   \\
 4  & 30  &           & 10.801   & 0.109   & -0.653  & 10.80  & 0.05  & -0.69   \\
 5  & 49  &           & 11.125   & 0.200   & -0.220  & 11.20  & 0.26  & -0.34   \\
\hline
\end{tabular}
\end{table*}

We have 5 stars in common with Feinstein et al (1976),
which are listed in Table~5. The mean differences turn out to be:

\[
V_{CP} - V_{FMF} = 0.021\pm0.054 
\]

\[
(B-V)_{CP} - (B-V)_{FMF} = 0.019\pm0.041 
\]

\[
(U-B)_{CP} -(U-B)_{FMF} = 0.073\pm 0.046   ,
\]

\noindent
where $CP$ indicates our photometry, and  $FMF$ stands for
Feinstein et al (1976).
Taking into account the different techniques used in extracting
the photometry, the agreement is very good both for magnitude and colors.\\

\begin{figure}
\centering
\includegraphics[width=9cm,height=9cm]{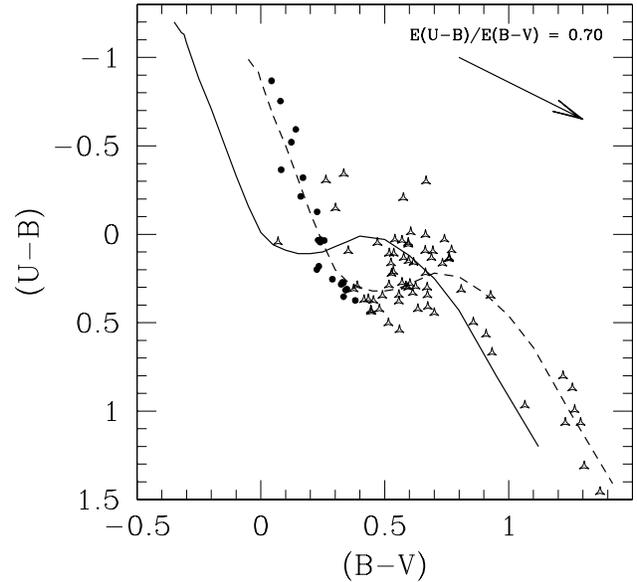}   
\caption{Two color diagram for the stars in the field of Collinder~228
brighter than $V$~=~17. The 
arrow indicates the reddening vector. The solid line is the empirical
un-reddened ZAMS from Schmidt-Kaler (1982), whereas the dashed  
line is  the same ZAMS, but  shifted by E$(B-V)$~=~0.30. See text for details.}
\end{figure}

\subsection{Reddening}
In order to obtain a rough estimate of the cluster mean reddening
we consider the distribution of the stars 
brighter than $V$~=~17 in the $(B-I)$ vs $(B-V)$ plane (see Fig.~8).
The selection in magnitude is done to limit the field stars 
contamination.
By applying the same technique described in Section~3.3,
we find that the bulk of stars have E$(B-V)$~=~0.40$\pm$0.20.
The uncertainty is due to the scatter 
of the stars in this plane, and indicates the presence
of stars with different reddening, in agreement with Feinstein
et al (1976) findings.\\
We use the $(U-B)-(B-V)$ diagram for all the stars brighter 
than $V$~=~17 to separate cluster candidate members (see Fig.~9).  The solid line
in this plot represents  the empirical
un-reddened ZAMS from Schmidt-Kaler (1982), whereas the dashed  
line is same ZAMS, but  shifted by E$(B-V)$~=~0.30.\\
\noindent
Two distinct populations seem to exist.
With filled circles we plotted all the stars having E$(B-V)$~=~0.30$\pm$0.05,
and we shall refer to them as to cluster candidate members.
A second population is defined by stars having  larger reddening
and is plotted with open triangles. All these stars are probably
just field stars.\\
These findings confirm  Feinstein et al (1976) results.\\

\begin{figure}
\centering
\includegraphics[width=9cm,height=9cm]{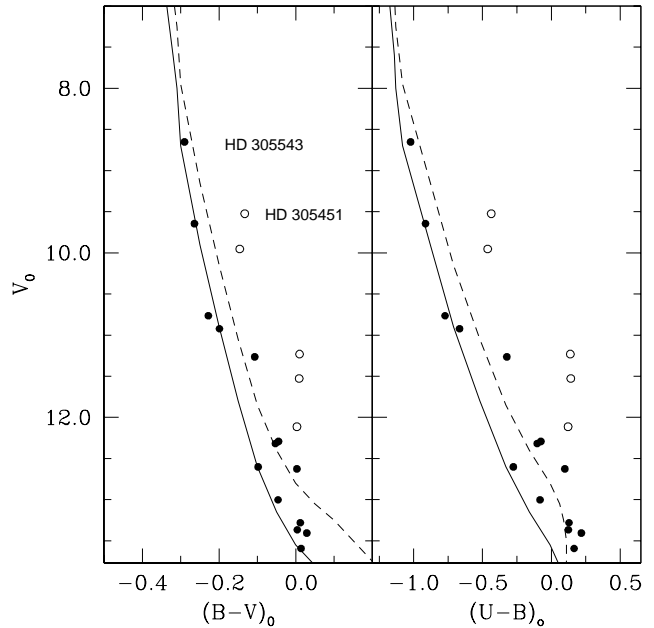}   
\caption{Reddening corrected CMDs for the 
candidate member stars in the field  of Collinder~228.
Over-imposed are solar abundance isochrones
for an age of 8$\times 10^{6}$~yrs. See text for details.}
\end{figure}

\begin{table*}
\tabcolsep 0.50cm
\caption{List of the new candidate members of Collinder~228 obtained in
 the present study.}
\begin{tabular}{cccccccc}
\hline
\hline
\multicolumn{1}{c}{ID} &
\multicolumn{1}{c}{$X$}  &
\multicolumn{1}{c}{$Y$}  &
\multicolumn{1}{c}{$V$} &
\multicolumn{1}{c}{$(B-V)$}  &
\multicolumn{1}{c}{$(U-B)$} &
\multicolumn{1}{c}{$(V-R)$} &
\multicolumn{1}{c}{$(R-I)$} \\
\hline
 4 &  -80.27 &  410.73  &  10.801 & 0.109 & -0.653 & 0.043 & 0.081 \\
 5 &   73.19 &  205.14  &  11.025 & 0.200 & -0.220 & 0.084 & 0.174 \\
 8 &  283.84 &  268.06  &  12.012 & 0.153 & -0.421 & 0.072 & 0.140 \\
 9 & -228.80 &  332.70  &  12.002 & 0.171 & -0.493 & 0.071 & 0.118 \\
12 & -249.81 &  193.19  &  12.275 & 0.293 & -0.206 & 0.154 & 0.198 \\
13 &  109.62 &  331.36  &  12.190 & 0.191 & -0.114 & 0.070 &-0.226 \\
14 & -386.07 &  261.02  &  12.187 & 0.319 &  0.354 & 0.100 & 0.138 \\
15 & -227.63 &   41.27  &  12.599 & 0.354 &  0.383 & 0.118 & 0.223 \\
18 &  268.82 &   85.25  &  13.269 & 0.270 &  0.145 & 0.160 & 0.227 \\
19 & -277.53 &  447.27  &  13.218 & 0.360 &  0.371 & 0.147 & 0.206 \\
20 &  484.68 &  411.66  &  13.369 & 0.286 &  0.135 & 0.168 & 0.214 \\
\hline
\end{tabular}
\end{table*}

\subsection{Age and distance}
In Fig.~10 we plot the reddening corrected CMDs for the stars of
 Collinder 228 
having E$(B-V)$~=~0.30$\pm$0.05 (i.e. the candidate members). 
They actually seem to form a tight sequence,
confirming our suggestion that they are good candidate members.
Over-imposed is
a theoretical solar metallicity isochrone (solid line)
from Girardi et al (2000) for an age of 8$\times 10^{6}$~yrs.
The same isochrone has been shifted by 0.75 mag (dashed line) to have an idea
of the MS broadening due to unresolved binaries. It is well known, in fact,
that binary stars define a sequence 0.75 mag brighter than the single
stars MS.
This permits us to suggest that five stars (indicated by open circles)
are probably non members, and that the four stars which lie close to the
binary sequence are probably unresolved binaries.\\
All the stars fainter than $V$~=~10.5 in these plots have not been
measured by Feinstein et al (1976), and hence we provide 11 new 
candidate members.
As a by-product, we infer an apparent distance modulus (m$-$M)~=~12.55$\pm$0.25,
which, once corrected for extinction, provides a distance of 1.9$\pm$0.2 kpc,
in agreement both with Feinstein et al (1976) and with Tapia et al (1988).\\
This corroborates the conclusion that Collinder~228 is closer to us than the
Carina nebula complex.\\

\noindent
In conclusion, in the observed region we identified 14 members, 3 in common
with Feinstein et al (1976) and 11 new, whose properties are
summarized in Table~6. Out of these,
4 are probably binaries.
Unfortunately, there is no overlap between our suggested binaries and
the study of Levato et al (1990). Only two stars are in common:
HD~305543, which is a member  and HD~305451, which probably is a field
star.
Nevertheless our result,  which comes from 
photometry,  although based on a small sample 
confirms their spectroscopic investigation that about 30\%
of the member stars are probably unresolved binaries.

\begin{figure}
\centering
\includegraphics[width=9cm,height=9cm]{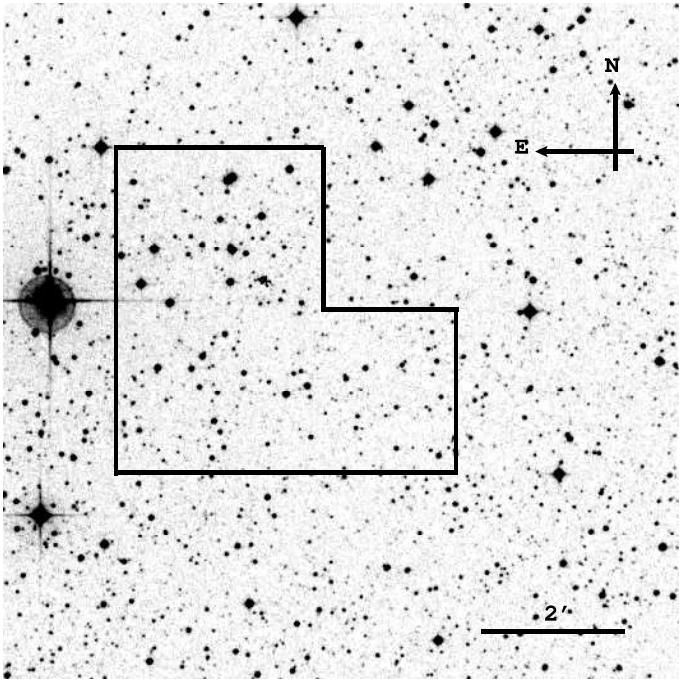}      
\caption{DSS map of a region around vdB-Hagen~99. The box confines the field
covered by our photometry.}
\label{h99}
\end{figure}

\section{The open cluster vdB-Hagen~99}
During a blue-red photographic survey of the southern Milky Way
van den Bergh \& Hagen (1975) provided a list of 262 known
or suspected open clusters. Among them, 64 groups were newly
recognized. One of this is the
scarcely populated and loose open cluster vdB-Hagen~99, which lies in the 
outskirts of the Carina complex. 

\subsection{Previous investigation}
vdB-Hagen~99 was studied by 
Landolt et al (1990), who emphasize the importance of this cluster
due to the probable membership of four known or suspected variables.
They obtained multicolor broad-band 
$UBVRI$ photoelectric
photometry for 48  stars, and intermediate- and narrow-band
photometry for 56 stars up to $V$~=~12. Moreover they obtained spectra for 21 stars
in the region of the cluster and additional photometry for 11 fainter stars,
with $ 13 \leq\ V \leq\ 16$.
The main results of their investigation are that vdB-Hagen~99
is a sparse open cluster with at least 24 candidate members.
The real existence  of the cluster is argued on the basis of the narrow
sequences the bright stars form in different color-color diagrams.
Although dominated by variable extinction, 
vdB-Hagen~99 has a low mean reddening 
E$(B-V)$~=~0.05. Moreover it is  $10^{8}$~yrs old and at a distance of 
about half a kpc. Finally it contains 8 photometric variables.

\begin{figure}
\centering
\includegraphics[width=9cm,height=9cm]{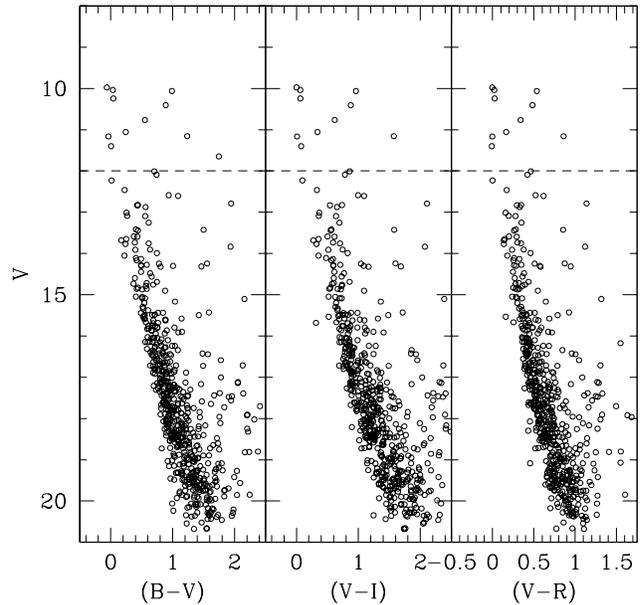}     
\caption{CMDs for all stars in the region of vdB-Hagen~99.
The dashed line indicates the limiting magnitude reached by 
Landolt et al (1990).}
\label{cmd_99}
\end{figure}

\begin{table*}
\tabcolsep 0.20cm
\caption{Photometry of the stars in the field
of  the open cluster vdB-Hagen~99 in common with Landolt et al (1990).
The name of the stars follow Landolt et al (1990). The suffix
$CP$ refers to the present study, whereas $LPLM$ indicates
Landolt et al (1990) photometry.}
\begin{tabular}{cccccccc}
\hline
\hline
\multicolumn{1}{c}{ID} &
\multicolumn{1}{c}{Name}  &
\multicolumn{1}{c}{$V_{CP}$}  &
\multicolumn{1}{c}{$(B-V)_{CP}$} &
\multicolumn{1}{c}{$(U-B)_{CP}$} &
\multicolumn{1}{c}{$V_{LPLM}$}  &
\multicolumn{1}{c}{$(B-V)_{LPLM}$} &
\multicolumn{1}{c}{$(U-B)_{LPLM}$} \\
\hline
 2  & CPD~-58~2451  & 10.037   & 0.086   & 0.047  & 10.078 & 0.114 & 0.093  \\
 4  & CPD~-58~2452  & 10.062   & 0.990   & 0.788  & 10.101 & 1.068 & 0.803  \\
 6  & CPD~-58~2447  & 10.765   & 0.553   & 0.127  & 10.782 & 0.648 & 0.175  \\
 7  & CPD~-58~2442  & 11.056   & 0.283   & 0.024  & 11.077 & 0.333 & 0.071  \\
 8  & CPD~-58~2440  & 10.244   & 0.094   & 0.012  & 10.236 & 0.127 & 0.085  \\
11  & VV~Car        & 11.748   & 1.747   & 1.319  & 11.775 & 1.784 & 1.356  \\
\hline
\end{tabular}
\end{table*}

\subsection{The present work}
We obtained CCD $UBVRI$ photometry for 900 stars in the region
shown in Fig.~\ref{h99}, up to $V$~=~20. Our survey supersedes
the previous one, whose limiting magnitude was about $V$~=~12.\\
The measured stars are shown in Fig.~12, in the planes
$V-(B-V)$, $V-(V-I)$ and $V-(V-R)$.
These CMDs resemble those  of NGC~3114 (see Fig.~2) and
Collinder~228 (see Fig.~7),
with a MS extending from $V$~=~10 up to $V$~=~20 and 
with some  evidence of the RG branch of the field stars population.
The similarity is not surprising, since all the clusters are projected
toward the Carina spiral arm.\\

\noindent
We have 6 stars in common with Landolt et al (1990),
which are listed in Table~7. The mean differences turn out to be:

\[
V_{CP} - V_{LPLM} = -0.023\pm0.016 
\]

\[
(B-V)_{CP} - (B-V)_{LPLM} = -0.053\pm0.025 
\]

\[
(U-B)_{CP} -(U-B)_{LPLM} = -0.044\pm 0.017   ,
\]

\noindent
where $CP$ indicates our photometry, whereas $LPLM$ stands for
Landolt et al (1990).
Taking into account the different techniques used in extracting
the photometry, the agreement is very good both for magnitude and colors.\\

\begin{figure}
\centering                                                                     \includegraphics[width=9cm,height=9cm]{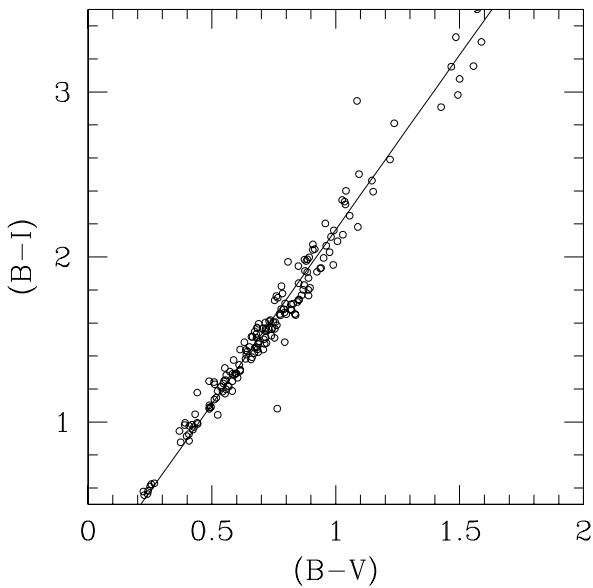}       
\caption{vdB-Hagen~99 stars brighter than $V$~=17 in the
$(B-I)-(B-V)$ plane.}
\label{bi99}
\end{figure}

\begin{figure}
\centering                                                                     \includegraphics[width=9cm,height=9cm]{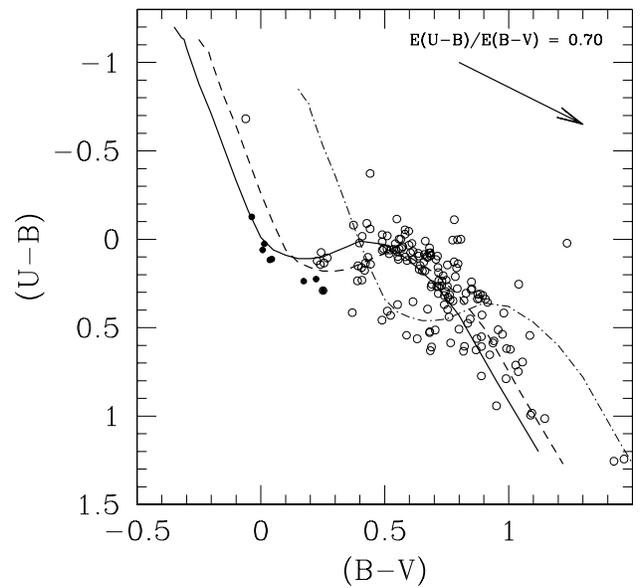}      
\caption{Two color diagram for the stars in the field of vdB-Hagen~99
 brighter than $V$~=~17. The 
arrow indicates the reddening vector. The solid line is the empirical
un-reddened ZAMS from Schmidt-Kaler (1982), while the dashed and dashed-dotted lines
are the same ZAMS, but  shifted by E$(B-V)$~=~0.10 and
E$(B-V)$~=~0.50, respectively.}
\label{two99}
\end{figure}

\subsection{Reddening}
To have an idea of the cluster mean reddening we selected all the stars
brighter than $V$~=17, and use their position  in 
$(B-I)-(B-V)$ plane, following the method described in Section~3.3.
The least squares fit yields E$(B-V)$~=~0.10$\pm$0.08. Again, 
the large uncertainty
is due to the scatter of the stars in this plane, and indicates the presence
of stars with different reddening, as already argued by 
Landolt et al (1990) .\\

\noindent
Candidate members can be searched for by considering the color color diagram
in Fig.~14, where filled circles represent stars having  
E$(B-V)$~=~0.04$\pm$0.03, whereas open circles indicate stars having
larger reddening. 
Tentatively, we suggest the possibility that two distinct populations
are actually present:
a group of eight
bright stars which have the same low reddening (filled circles), and 
all the other stars
which have a larger reddening with a significant scatter.\\
We argue that the brighter stars are candidate members of vdB-Hagen~99,
whereas all the other stars having larger value of E$(B-V)$ (see Fig~14)
are probably field stars.

\subsection{Distance and age}
In order to test this hypothesis, we construct the
reddening corrected CMDs in the $V_o~-~(B-V)_o$ and $V_o~-~(U-B)_o$ planes  
for all the stars for which we could obtain a 
reddening estimate (see Fig.~15).
Filled symbols indicate cluster candidate members, whereas open symbols 
indicated background stars.\\
As suggested above, two distinct populations are readily visible. \\
Most of the stars
we measured are located beyond vdB-Hagen~99, at the distance of the Carina 
spiral arm (2.5-3.0 kpc). They are indicated with open symbols, 
and fitted with an 
empirical ZAMS shifted by $(m-M)_o$~=~12.20. \\
With filled triangles we indicate Landolt et al (1990) candidate members,
20 stars in total. 
They define a tight sequence along the empirical ZAMS (solid line)
shifted by $(m-M)_o$~=~8.30.
Noticeably, all the empty triangles, which identify stars that have been suggested 
by Landolt et al (1990)
not be cluster members, actually lie close to the field stars sequence.\\
The stars indicated with filled circles are probable cluster members observed by us.
Four of them - namely  CPD~-58~2451, CPD~-58~2440, CPD~-58~2442 and CPD~-58~2447 - 
are in common with Landolt et al (1990). The other 4 are probable new
candidate members, and their properties are summarized in Table~.8.
The remaining two common stars (CPD~-58~2452 and VV~Car) are red stars of $GK$
spectral type belonging to the field.\\
This way we increased the number of cluster members, suggesting that
they are at least 28.
Finally, the stars redder than vdB-Hagen~99 members are probably interlopers
stars, located between us and the cluster.\\
As for the age, most of the stars lie close to the ZAMS, 
with the exception of the brightest
ones. This is an indication that the cluster is young, as
already claimed by  Landolt et al (1990). To have an idea of the cluster age,
we over-imposed in Fig.~15 a solar metallicity isochrone (dotted line)
from Girardi et al (2000)
for the age of $1.25 \times 10^{8}$ yrs, which nicely fits the evolved
stars.\\

In conclusion,
these results confirm that vdB-Hagen~99 is a young cluster projected toward
the Carina spiral arm, at a distance of about 500 pc from the Sun.
The mean reddening of cluster members turns out to be E$(B-V)$~=~0.04$\pm$0.03,
in agreement with the findings of Landolt et al (1990).

\begin{figure}
\centering
\includegraphics[width=9cm,height=9cm]{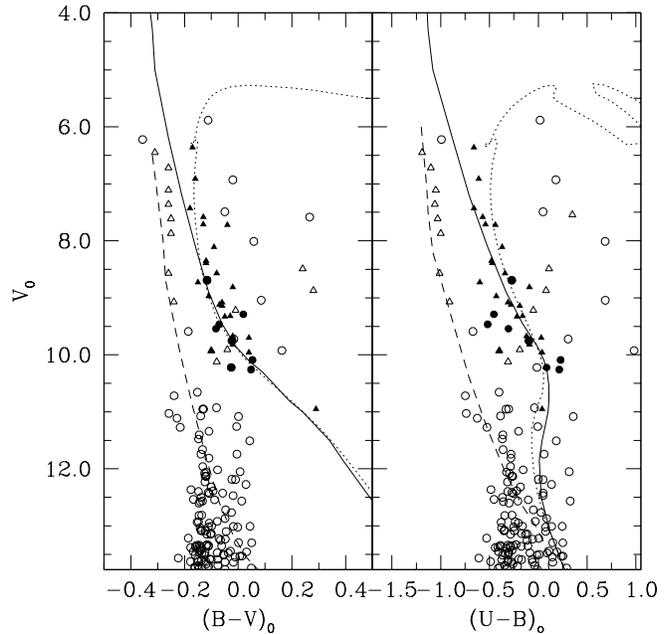}     
\caption{Reddening corrected CMD for the stars in the field of vdB-Hagen~99.
Filled symbols indicate cluster members, 
whereas open circles are field stars. Triangles refer to Landolt et al (1990)
photometry, whilst circles indicate stars whose photometry has been presented in this study. Finally, the dotted line is an isochrone
for the age of $1.25 \times 10^{8}$ yrs, whereas the other two lines
are the same empirical ZAMS, shifted by (m$-$M)~=12.20 (dashed line)
and (m$-$M)~=8.30 (solid line). See text for additional details.}
\label{iso99}
\end{figure}

\section{Conclusions}
In this paper we presented $UBVRI$ CCD photometry
for three young clusters in the direction of the Carina
spiral feature: NGC~3114, Collinder~228 and vdB-Hagen~99.
Our results can be summarized as follows:

\begin{description}
\item $\bullet$ NGC~3114 lies
in the outskirts of the Carina nebula, at a distance of 920 pc from the Sun.
In the field we studied, we isolate 7 cluster members, which has a 
low reddening (E$(B-V)$~=~0.07$\pm$0.03) and a maximum age of $3 \times 10^{8}$~yrs. \\
\item $\bullet$ Collinder~228 is a younger (8$\times 10^{6}$) yrs cluster located
in front of the Carina nebula complex. We identify 11 new candidate members
and confirm that 30\% of the cluster members are probably unresolved
binaries.\\
\item $\bullet$ Finally, vdB-Hagen~99 is a loose open cluster about
 $10^{8}$ yrs old, and  with a fairly low mean reddening. We confirm previous results,
and add 4 new members.
\end{description}

\noindent
Putting together the  results of this paper and those of  
previous works (Patat \&
Carraro 2001; Carraro et al 2001),  the Carina complex turns out
to be populated by very young clusters (like Collinder~228, NGC~3324 and
Bochum~11)
in its inner region, and by young or
intermediate age objects (like NGC~3114, vdB-Hagen~99 and Loden~165)
in its outskirts. \\
\noindent
This confirms previous suggestions by Feinstein (1995) 
about the existence of an age gradient in the Carina complex.
We shall discuss this issue in more details in forthcoming papers,
when all the clusters observed in our survey will be analysed.\\

\begin{table*}
\tabcolsep 0.50cm
\caption{List of the new candidate members of vdB-Hagen~99 obtained in
 the present study.}
\begin{tabular}{cccccccc}
\hline
\hline
\multicolumn{1}{c}{ID} &
\multicolumn{1}{c}{$X$}  &
\multicolumn{1}{c}{$Y$}  &
\multicolumn{1}{c}{$V$} &
\multicolumn{1}{c}{$(B-V)$}  &
\multicolumn{1}{c}{$(U-B)$} &
\multicolumn{1}{c}{$(V-R)$} &
\multicolumn{1}{c}{$(R-I)$} \\
\hline
 3  & 182.75 &  235.10 & 10.244 &  0.044 &  0.212 &  0.026 & 0.027 \\
 8  & 106.44 &  341.37 & 11.165 & -0.037 & -0.027 & -0.005 & 0.005 \\
10  &  -7.76 & -166.28 & 11.399 &  0.007 &  0.160 &  0.008 & 0.080 \\
13  & 242.58 &  -76.50 & 12.229 &  0.014 &  0.126 &  0.000 & 0.081 \\
\hline
\end{tabular}
\end{table*}

\begin{acknowledgements}
GC acknowledges kind hospitality from ESO.
This study made use of Simbad and WEBDA. The suggestions of an anonymous referee
are warmly acknowledged.
\end{acknowledgements}

{}

\end{document}